\documentclass[12pt,a4paper,reqno]{article}
\usepackage{amsmath}\usepackage{amssymb}
\usepackage{amsthm}

\newcommand{\upline}{\vspace{-\abovedisplayskip}\vspace{-\baselineskip}}
\newcommand{\be}{\begin{equation}}
\newcommand{\ee}{\end{equation}}
\renewcommand{\(}{\left(}
\renewcommand{\)}{\right)}

\newcommand{\e}{\mathrm{e}}
\renewcommand{\i}{\mathrm{i}}
\renewcommand{\t}{\mathbf{t}}

\newcommand{\R}{\mathbb{R}}

\newcommand{\ox}{\otimes}
\newcommand{\<}{\langle}
\renewcommand{\>}{\rangle}
\newcommand{\half}{\tfrac{1}{2}}
\newcommand{\hlf}[1]{\tfrac{#1}{2}}

\newcommand{\quarter}{\tfrac{1}{4}}
\newcommand{\qqquad}{\quad\quad\quad}

\newcommand{\tr}{\operatorname{tr}}

\theoremstyle{plain} 

\theoremstyle{definition}

\theoremstyle{remark}

\numberwithin{equation}{section}
\begin{document} \title{On local invariants of pure three-qubit states}

\author{\\\\Anthony Sudbery 
\\\\\small Department of Mathematics
\\\small University of York 
\\\small Heslington 
\\\small York 
\\\small England YO1 5DD 
\\\small  Email:  as2@york.ac.uk\\\\}  
\date{24 February 2000, revised 15 November 2000} 
\maketitle 
\vspace{2cm} 

\begin{abstract} We study invariants of three-qubit states under local
unitary transformations, i.e.\ functions on the space of entanglement
types, which is known to have dimension 6. We show that there is no set
of six algebraically independent polynomial invariants of degree $\le 6$, and find
such a set with maximum degree 8. We describe an intrinsic definition
of a canonical state on each orbit, and discuss the (non-polynomial)
invariants associated with it. \end{abstract}

\newpage
\section{Introduction} \label{S:Intro}

The invariants of many-particle states under unitary transformations
which act on single particles separately (``local" transformations) are
of interest \cite{HilaryNSme, Grasslinv, NoahSandu, NoahSandume,
 Rains} because they give the finest discrimination between
different types of entanglement. They can be regarded as coordinates on
the space of entanglement types (equivalently, the space of orbits of
the group of local transformations). In this paper we study the case of
pure states of three spin-$\half$ particles, or qubits. For mixed states
of two qubits, it is possible to give a complete set of invariants
\cite{Makhlin}, describing the 9-dimensional space of orbits in terms
of 18 invariants, nine of which may be taken to have only discrete values (for example,
the signs of certain polynomials listed in \cite{Makhlin}). For pure three-qubit states,
where the space of orbits is known \cite{HilaryNSme} to be
6-dimensional, we can at present do no more than find a set of six
algebraically independent invariants. We will show (Section 3) that in order to do
this with polynomials in the state coordinates it is necessary to go to
polynomials of order 8, and we will exhibit (Section 4) a set of six
independent invariants; their physical meaning is discussed in Section 5. We will
also discuss (Section 6) the possibility of finding a more convenient
set of non-polynomial invariants. Section 2 is an introductory
discussion of the invariants of pure $n$-qubit states.

\section{Pure states: general considerations}

A general theory of local invariants of mixed $n$-particle states has
been given by Rains and Grassl et al.\ \cite{Grasslinv, Rains}. Here we review the part of that theory that refers to pure states.

The most general system is that of $n$ non-identical particles $A, B,
\ldots$ with one-particle state spaces of dimensions $d_A, d_B,\ldots$. 
Let $\{|\psi_i^X\> : i = 1,\ldots , d_X\}$ be an orthonormal
basis of one-particle states of particle $X$; then the general
$n$-particle state can be written 
\[ |\Psi\> = \sum_{ijk\cdots}
t^{ijk\cdots}|\psi_i^{(A)}\>|\psi_j^{(B)}\>|\psi_k^{(C)}\>\cdots
\]
where the sum is over values of $i$ from 1 to $d_A$, values of $j$ from
1 to $d_B$, and so on. By the First Fundamental Theorem of invariant
theory \cite{Weyl:CG} applied to $U(d_A), U(d_B),\ldots$, any polynomial
in $t^{ijk\cdots}$ which is invariant
under the action on $|\Psi\>$ of the local group
$U(d_A)\times U(d_B)\times\cdots$ is a sum of homogeneous polynomials of
even degree (say $2r$), of the form 
\be \label{eq:standard}
  P_{\sigma\tau\cdots }(\t )=t^{i_1 j_1 k_1 \cdots }\cdots t^{i_r j_r
  k_r\cdots}\overline{t}_{i_1 j_{\sigma (1)} k_{\tau
  (1)}\cdots}\cdots\overline{t}_{i_r j_{\sigma (r)} k_{\tau (r)}\cdots}
\end{equation}
where $\sigma, \tau, \ldots$ are permutations of $(1,\ldots ,r)$. Here
$\overline{t}_{ijk\cdots}$ is the complex conjugate of $t^{ijk\cdots}$,
and we adopt the usual summation convention on repeated indices, one in
the upper position and one in the lower. Note that $P_{\sigma \tau
\cdots}$ is unchanged by simultaneous conjugation of the permutations
$\sigma, \tau , \ldots$:
\[ P_{\sigma\tau\cdots}(\t ) = P_{\sigma^\prime \tau^\prime \cdots} (\t )
\quad \text{ if } \quad \sigma^\prime = \kappa\sigma\kappa^{-1}, \quad
\tau^\prime = \kappa\tau\kappa^{-1}, \quad \ldots
\]
since such a conjugation merely expresses the effect of changing the
order of the factors in each summand in $P$.

For two particles $A,B$ there is just one permutation $\sigma$, which we can decompose
into cycles $\kappa_1 ,\ldots ,\kappa_s$ of orders $l_1,\ldots ,l_s$
with $l_1 +\cdots +l_s = r$. The polynomial $P_\sigma (\t )$ then splits
into a product of polynomials $P_{\kappa_1}\cdots P_{\kappa_s}$, where
$P_\kappa$ depends only on the order of the cycle $\kappa$, which is
equal to half the degree of $P_\kappa$:
\begin{align*}
P_\kappa (\t ) &= t^{i_1 j_1}\overline{t}_{i_1 j_{\kappa (1)}}
t^{i_{\kappa(1)}j_{\kappa (1)}}\overline{t}_{i_{\kappa (1)}j_{\kappa^2(1)}}
 \cdots \\ 
&= t^{i_1 j_1}\overline{t}_{i_1 j_2} t^{i_2 j_2}\overline{t}_{i_2 j_3}
\cdots t^{i_l j_l}\overline{t}_{i_l j_1}
\end{align*}
(by renaming the dummy indices $j_{\kappa (1)}, j_{\kappa^2(1)}, \ldots
, j_{\kappa^{l-1}(1)}$)
\[ = \tr (\rho_B^l)
\]
where $\rho_B = \tr_A |\Psi\>\<\Psi|$ is the density matrix of particle
$B$, with matrix elements
\[ (\rho_B)^j_k = t^{ij}\overline{t}_{ik}.
\]

Thus the polynomial invariants of a two-particle pure state are the sums
of the powers of the eigenvalues of $\rho_B$. These can all be expressed
in terms of the first $d_B$ power-sums, which generate the algebra of
invariant polynomials and are algebraically independent if the eigenvalues are
independent. However, they are not independent if $d_A < d_B$, for in
that case some of the eigenvalues of $\rho$ vanish. But clearly the
same argument could be used to show that the algebra of invariants is
generated by the traces of the powers of $\rho_A$, which is consistent
because the non-zero eigenvalues of $\rho_A$ are the same as those of
$\rho_B$. Thus the algebra of polynomial invariants of two-particle pure
states has a set of independent generators 
\[ \tr (\rho_A^l) = \tr (\rho_B^l), \qqquad l=1,\ldots ,\min(d_A, d_B).
\]

The non-zero eigenvalues of $\rho_A$ (or $\rho_B$) are in fact the
squares of the coefficients in the Schmidt decomposition of $|\Psi\>$,
so what we have here is the well-known fact that the local invariants
of a pure two-particle state are the symmetric functions of the
Schmidt coefficients.

\section{Polynomial invariants of three-qubit states}

For the remainder of the paper we consider three spin-$\half$ particles
$A,B,C$. The classification of pure states of this system has been
discussed in \cite{Hilaryme, SchlienzMahler}, and their  invariants in
\cite{Woottang, Kempeinv}. It is known \cite{NoahSandu} that
the dimension of the space of orbits is 6; there are therefore six
algebraically independent local invariants. We will show that there are
no more than five algebraically independent invariants of degree less
than 8, and exhibit a set of six algebraically independent invariants
with maximum degree 8.\footnote{I understand that similar conclusions
have been reached by Markus Grassl \cite{Grassltalk}.}

The vector space of homogeneous invariants of degree $2r$ is spanned by functions
$P_{\sigma\tau}$ labelled by pairs of elements of $S_r$, the group of
permutations of $r$ things. Thus there is one independent invariant of
degree 2, 
\[I_1 = P_{ee}(\t ) = t^{ijk}\overline{t}_{ijk} = \<\Psi|\Psi\>
\] 
where $e$ is the identity permutation, so that $S_1=\{e\}$. If
$S_2=\{e,\sigma\}$, the four linearly independent quartic invariants are
\begin{align*} P_{ee}(\t ) &= t^{i_1 j_1 k_1}\overline{t}_{i_1 j_1 k_1}
t^{i_2 j_2 k_2}   \overline{t}_{i_2 j_2 k_2} = \<\Psi |\Psi\>^2, \\ I_2
= P_{e\sigma}(\t ) &= t^{i_1 j_1 k_1}\overline{t}_{i_1 j_1 k_2}  t^{i_2
j_2 k_2}\overline{t}_{i_2 j_2 k_1} = \tr (\rho_C^2), \\ I_3 = P_{\sigma
e}(\t ) &= t^{i_1 j_1 k_1}\overline{t}_{i_1 j_2 k_1}  t^{i_2 j_2
k_2}\overline{t}_{i_2 j_1 k_2} = \tr (\rho_B^2), \\ I_4 = P_{\sigma
\sigma}(\t ) &= t^{i_1 j_1 k_1}\overline{t}_{i_1 j_2 k_2} t^{i_2 j_2
k_2}\overline{t}_{i_2 j_1 k_1} = \tr (\rho_A^2) 
\end{align*} 
where
$\rho_A, \rho_B, \rho_C$ are the one-particle density matrices: 
\[
\rho_X = \tr_{YZ}|\Psi\>\<\Psi| \text{ where }  \{X,Y,Z\} = \{A,B,C\}
\text{ in some order.} 
\] 
Thus there are at most four algebraically
independent invariants of degree $\le 4$.

Higher-order invariants $P_{\pi\sigma}(\t )$ with $\pi ,\sigma\in S_3$ are
functions of the four quadratic and quartic invariants if $\pi$ and
$\sigma$ are equal or if either of them is the identity. To see this,
note first that if $\pi = \sigma$,
\begin{align*} 
P_{\sigma\sigma}(\t ) &= t^{i_1 j_1 k_1} \cdots t^{i_r j_r k_r} \overline{t}_{i_1
j_{\sigma (1)}k_{\sigma (1)}} \cdots\overline{t}_{i_r j_{\sigma (r)}k_{\sigma(r)}} \\
&= (\rho_A)^{i_1}_{i_{\tau(1)}}(\rho_A)^{i_2}_{i_{\tau (2)}}\cdots
(\rho_A)^{i_r}_{i_{\tau(r)}}
\end{align*}
where $\tau = \sigma^{-1}$. This is a product of traces of powers of
$\rho_A$. But since $\rho_A$ is a $2\times 2$ matrix, the
Cayley-Hamilton theorem enables us to express $\tr (\rho_A^r)$ for $r\ge
3$ as a function of $\tr\rho_A$ and $\tr\rho_A^2$.

Secondly, if $\pi = e$, 
\begin{align*}
P_{e\sigma}(\t ) &= t^{i_1 j_1 k_1}\cdots t^{i_r j_r k_r}\overline{t}_{i_1
j_1 k_{\sigma (1)}} \cdots \overline {t}_{i_r j_r k_{\sigma (r)}} \\
&= (\rho_C)^{k_1}_{k_{\sigma (1)}}\cdots (\rho_C)^{k_r}_{k_{\sigma (r)}}
\end{align*}
which is a product of traces of powers of $\rho_C$; and similarly
$P_{\pi e}(\t )$ is a product of traces of powers of $\rho_B$.

Thus the only sextic invariants $P_{\pi\sigma}$
which might be algebraically independent of the quadratic and quartic invariants are those
for which $\pi$ and $\sigma$ are distinct 2-cycles, or distinct
3-cycles, or one is a 2-cycle and the other is a 3-cycle. Moreover, in
each of these categories all the possible pairs $(\pi, \sigma)$ are related by
simultaneous conjugation and therefore give the same invariant. There
are therefore three possible independent sextic invariants:

1. $\pi$, $\sigma$ distinct 3-cycles, say $\pi = (123)$, $\sigma =
(132)$. This gives
\begin{align} I_5 = P_{(123)(132)}(\t ) &= t^{i_1 j_1 k_1}t^{i_2 j_2 k_2}
t^{i_3 j_3 k_3}\overline{t}_{i_1 j_2 k_3}\overline{t}_{i_2 j_3 k_1}
\overline{t}_{i_3 j_1 k_2} \notag \\
&= (\rho_{BC})^{j_1 k_1}_{j_2 k_3}(\rho_{BC})^{j_2 k_2}_{j_3 k_1}
(\rho_{BC})^{j_3 k_3}_{j_1 k_2}
\end{align}
where $\rho_{BC}=\tr_A|\Psi\>\<\Psi|$ is the density matrix of the
two-particle system of particles $B$ and $C$. This invariant was identified
by Kempe \cite{Kempeinv} as one which distinguishes three-particle
states which have identical density matrices for every subsystem. It has
exactly the same form when expressed as a function of $\rho_{AB}$ or of
$\rho_{AC}$. 

2. $\pi$, $\sigma$ distinct 2-cycles, say $\pi = (12)$, $\sigma = (23)$. This
gives
\begin{align}   I_5^\prime = P_{(12)(23)}(\t ) 
&= t^{i_1 j_1 k_1}t^{i_2 j_2 k_2}t^{i_3 j_3k_3}\overline{t}_{i_1 j_2 k_1}
\overline{t}_{i_2 j_1k_3}\overline{t}_{i_3 j_3 k_2} \notag \\
&= (\rho_B )^{j_1}_{j_2}(\rho_C)^{k_3}_{k_2}
(\rho_{BC})^{j_2 k_2}_{j_1 k_3} \notag \\
&= \tr [(\rho_B\ox \rho_C) \rho_{BC}].
\end{align}

3. $\pi$ a 2-cycle, say (12), and $\sigma$ a 3-cycle, say (123), or
vice versa. These give
\begin{align}   I_5^{\prime\prime} = P_{(12)(123)}(\t ) 
&= t^{i_1 j_1 k_1}t^{i_2 j_2 k_2}t^{i_3 j_3k_3}\overline{t}_{i_1 j_2 k_2}
\overline{t}_{i_2 j_1 k_3}\overline{t}_{i_3 j_3 k_1} \notag \\
&= (\rho_{AC})^{i_1 k_1}_{i_2 k_3}(\rho_A)^{i_2}_{i_1} (\rho_C)^{k_3}_{k_1}
\notag \\
&= \tr [(\rho_A\ox \rho_C)\rho_{AC}]
\end{align}
and
\begin{align}   I_5^{\prime\prime\prime} = P_{(123)(12)}(\t ) 
&= t^{i_1 j_1 k_1}t^{i_2 j_2 k_2}t^{i_3 j_3 k_3}\overline{t}_{i_1 j_2 k_2}
\overline{t}_{i_2 j_3 k_1}\overline{t}_{i_3 j_1 k_3} \notag \\
&= \tr [(\rho_A\ox\rho_B)\rho_{AB}].
\end{align}

Primes have been placed on the symbols for these last three invariants
because they will not feature in our final list of independent
invariants, each of them being expressible in terms of $I_5$ and the
quadratic and quartic invariants. To show this, we write $I_5$ in terms
of $2\times 2$ matrices by considering the $4 \times 4$ matrix
$\rho_{BC}$ as a set of four $2 \times 2$ matrices $X^{j_1}_{j_2}$: the
matrix elements of $X^{j_1}_{j_2}$, labelled by $(k_1, k_2)$, are
\[ (X^{j_1}_{j_2})^{k_1}_{k_2} = (\rho_{BC})^{j_1 k_1}_{j_2 k_2}.
\]
Then
\[ I_5 = \tr (X^{j_1}_{j_2} X^{j_3}_{j_1} X^{j_2}_{j_3}).
\]
Now we use the $2 \times 2$ matrix identity
\be \label{trilinid} \begin{split} \tr (XYZ) + \tr (XZY) =
\tr X \tr (YZ) &+ \tr Y \tr (ZX) + \tr Z \tr (XY)\\
 &- \tr X \tr Y \tr Z \end{split}
\end{equation}
which holds for any $2 \times 2$ matrices $X,Y,Z$, and can be obtained
by trilinearising (or ``polarising" \cite{Weyl:CG} --- replace $X$ first by
$X+Y$ and then by $X+Y+Z$) the cubic identity
\[ \tr X^3 = \hlf{3}\tr X \tr X^2 - \half (\tr X)^3
\] 
which in turn is obtained by taking the trace of the Cayley-Hamilton
theorem. Apply \eqref{trilinid}  to the matrices $X^{j_1}_{j_2}$,
$X^{j_3}_{j_1}$, $X^{j_2}_{j_3}$ occurring in the expression for $I_5$.
The first term on the left-hand side is $I_5$; the second is 
\[ \tr (X^{j_1}_{j_2}X^{j_2}_{j_3}X^{j_3}_{j_1}) = \tr (\rho^3_{BC})
= \tr(\rho_A^3)
\]
since the non-zero eigenvalues of $\rho_{BC}$ are the same as those of
$\rho_A$ (both being the squares of the coefficients in a Schmidt
decomposition of $|\Psi\>$). The first term on the right-hand side is 
\begin{align*} \tr (X^{j_1}_{j_2})\tr (X^{j_3}_{j_1}X^{j_2}_{j_3}) 
&= (\rho_B)^{j_1}_{j_2}(\rho_{BC})^{j_3 k_1}_{j_1 k_2}
(\rho_{BC})^{j_2 k_2}_{j_3 k_1} \\
&= (\rho_B)^{j_1}_{j_2}t^{i_1 j_3 k_1}\overline{t}_{i_1 j_1 k_2}
t^{i_2 j_2 k_2}\overline{t}_{i_2 j_3 k_1} \\
&= (\rho_B)^{j_1}_{j_2}(\rho_A)^{i_1}_{i_2}(\rho_{AB})^{i_2 j_2}_{i_1 j_1} \\
&= \tr [(\rho_A\ox \rho_B)\rho_{AB}];
\end{align*}
the second and third terms differ from the first only by permuting the
indices $j_1, j_2, j_3$ and therefore (after summing) are equal to it;
and the last term is
\[ \tr (X^{j_1}_{j_2})\tr (X^{j_2}_{j_3})\tr(X^{j_3}_{j_1}) = \tr
(\rho_B^3).
\]
Thus \eqref{trilinid} gives
\be \label{eq:I5}
I_5 = 3\tr[(\rho_A\ox \rho_B)\rho_{AB}] - \tr (\rho_A^3) - \tr (\rho_B^3).
\end{equation}
Similarly, using the alternative expressions for $I_5$ in terms of
$\rho_{AB}$ and $\rho_{AC}$ gives
\begin{align}
I_5 &= 3\tr[(\rho_B\ox\rho_C)\rho_{BC}]-\tr (\rho_B^3)-\tr (\rho_C^3)
\label{eq:I5'}\\
&= 3\tr[(\rho_A\ox \rho_C) \rho_{AC}] - \tr (\rho_A^3) - \tr (\rho_C^3).
\label{eq:I5"}\end{align}

So there are at most five independent invariants of degree 6 or less.
Since six invariants are needed to parametrise the orbits
\cite{NoahSandu}, we must use at least one invariant of degree 8 or
more. A convenient, and physically significant, choice is the 3-tangle
identified by Coffman, Kundu and Wootters \cite{Woottang}:
\be I_6 = \quarter\tau_{123}^2 =
\left|\epsilon_{i_1 i_2}\epsilon_{i_3 i_4}\epsilon_{j_1 j_2}\epsilon_{j_3 j_4}
\epsilon_{k_1 k_3}\epsilon_{k_2 k_4}t^{i_1 j_1 k_1}t^{i_2 j_2 k_2}
t^{i_3 j_3 k_3}t^{i_4 j_4 k_4}\right|^2
\end{equation}
where $\epsilon_{ij}$ is the antisymmetric tensor in two dimensions
($\epsilon_{12}=-\epsilon_{21}=1$, $\epsilon_{11}=\epsilon_{22}=0$). The
expression between the modulus signs is an SU$(2)^3$ invariant (though not a
U$(2)^3$ invariant --- its phase is not invariant under local
transformations), so its modulus is a local invariant. The invariant $I_6$
can be put into our standard form of a sum of terms like \eqref{eq:standard} by multiplying the
SU$(2)^3$ invariant by its complex conjugate
\[  \epsilon^{i_5 i_6}\epsilon^{i_7 i_8}\epsilon^{j_5 j_6}\epsilon^{j_7 j_8}
  \epsilon^{k_5 k_7}\epsilon^{k_6 k_8}
  \overline{t}_{i_5 j_5 k_5}\overline{t}_{i_6 j_6 k_6}
  \overline{t}_{i_7 j_7 k_7}\overline{t}_{i_8 j_8 k_8}
\]
(where the contravariant tensor $\epsilon^{ij}$ is numerically the same
as $\epsilon_{ij}$), and using the identity
\[\epsilon^{ab}\epsilon_{cd} = \delta^a_c\delta^b_d -\delta^a_d\delta^b_c.
\]

To show that the invariants $I_1,\ldots,I_6$ are independent it is
sufficient to show that their gradients are linearly independent at some
point. To calculate these gradients in the 16-(real)dimensional space of pure
states, we can treat $t^{ijk}$ and $\overline{t}_{ijk}$ formally as independent
coordinates; the fact that our invariants are real means that the 16 componentsof the gradient of $I_a$ are the real and imaginary parts of the partial derivatives with respect to $t^{ijk}$. The results of calculating $\partial I_a
/\partial t^{ijk}$ and putting 
\[ t^{000}=t^{010}=t^{110}=0, \quad
t^{011}=t^{100}=t^{101}=t^{111}=1, \quad t^{001}=\i,
\]
\[ \overline{t}^{ijk} = \text{  complex conjugate of  } t^{ijk}
\] 
(where $0$ and $1$ are the two possible values of $i,j,k$) are as
follows:
\begin{align*}
\partial_\t I_1 &= (0,-\i,0,1,1,1,0,1)\\
\partial_\t I_2 &= (-2\i,-8\i,2,8,4,10,2,8)\\
\partial_\t I_3 &= (0,2-8\i,0,6-2\i,6,8-2\i,2+2\i,6+2\i)\\
\partial_\t I_4 &= (2-2\i,2-6\i,0,6-2\i,6,8-2\i,0,8+2\i)\\
\partial_\t I_5 &= (6-9\i,12-36\i,6,30-12\i,21,45-12\i,9+6\i,36+12\i)\\
\partial_\t I_6 &= (-8,0,-8+16\i,8,8,0,-8\i,0)
\end{align*}
These six vectors are indeed linearly independent over $\R$.

\section{Physical significance of the invariants}

The invariant $I_1$ is just the norm of the three-party state and
therefore has no physical significance; we will normally set it equal to
1. The three invariants $I_1,I_2,I_3$ are one-particle quantities, giving the
eigenvalues of the one-particle density matrices; they are equivalent to
the one-particle entropies, which measure how entangled each particle is
with the other two together. The entanglement in each pair of particles
and the three-way entanglement of the whole system are all given by the
last invariant $I_6$, as follows. A good measure of the entanglement of
two qubits $A,B$ in a mixed state is the \emph{2-tangle} $\tau_{AB}$
\cite{Woottang}, which is a monotonic function of the entanglement of
formation \cite{Woot:entform}. The three-way entanglement of three
qubits $A,B,C$ in a pure state is measured by the \emph{3-tangle}
\[
\tau_{ABC} = \tau_{A(BC)} - \tau_{AB} - \tau_{AC}
\]
where $\tau_{A(BC)}=4\det\rho_A=2(I_1^2-I_2)$ is another measure
(equivalent to the entropy of $A$) of how entangled $A$ is with the pair
$(BC)$. It can be shown \cite{Woottang} that $\tau_{ABC}$ is invariant
under permutations of $A$, $B$ and $C$; in fact it is equal to our
invariant $I_6$. By solving the equations expressing the permutation
invariance of $\tau_{ABC}$, we can now give formulae for all three
2-tangles and the 3-tangles in terms of our invariants:
\begin{align*}
\tau_{AB}&= 1 - I_2 - I_3 + I_4 -\half I_6,\\
\tau_{AC}&= 1 - I_2 + I_3 - I_4 -\half I_6,\\
\tau_{BC}&= 1 + I_2 - I_3 - I_4 -\half I_6,
\end{align*}
\upline
\[
\tau_{ABC} = I_6.
\]
The 3-tangle $I_6$ is maximal for the GHZ state $|000\> + |111\>$, whose
2-tangles vanish; on the other hand, $I_6$
vanishes at the states $p|100\> + q|010\> + r|001\>$.

The remaining invariant, $I_5$, is a different and independent measure
of the entanglement of each pair of qubits. Its existence shows that the
2-tangles and 3-tangle are not sufficient to determine a pure 3-qubit
state up to local equivalence. As is shown by eqs.\ \eqref{eq:I5}--\eqref{eq:I5"}, this invariant is equivalent to any one of the
two-qubit quantities $\kappa_{AB}=\tr[(\rho_A\ox\rho_B)\rho_{AB}]$ 
(together with one-qubit quantities), and it relates these three 2-qubit quantities to
each other. If we regard the hermitian operators $\rho_A$ and $\rho_B$
as observables, then $\kappa_{AB}$ is the expectation value of
$\rho_A\rho_B$, so $\kappa_{AB} - I_2 I_3$ is the correlation between
the eigenvalues of $\rho_A$ and $\rho_B$. It is related to the relative
entropy of the two-qubit state $\rho_{AB}$ relative to the product state
$\rho_A\ox\rho_B$, and is a second measure of the entanglement of the
pair $(A,B)$, independent of the 2-tangle $\tau_{AB}$.

Finally, we give the values of these invariants for some special states
(all of which are taken to be normalised).

For a factorised state $a|111\> + b|100\>$,
\[
I_2 = 1, \quad I_3 = I_4 = a^4 + b^4, \quad I_5 = a^6 + b^6, \quad I_6 = 0.
\]
For a generalised GHZ state $p|000\> + q|111\>$,
\[
I_2 = I_3 = I_4 = p^4 + q^4, \quad I_5 = p^6 + q^6, \quad I_6 = 4p^2q^2.
\]
For the minimally 3-tangled \cite{Woottang} state $p|100\> + q|010\> +
r|001\>$,
\[
I_2 = p^4 + (q^2 + r^2)^2, \quad I_3 = q^4 + (r^2 + p^2)^2, \quad I_4 =
r^4 + (p^2 + q^2)^2,
\]
\upline
\[
I_5 = p^6 + q^6 + r^6 + 3p^2q^2r^2, \quad I_6 = 0.
\]

\section{Canonical coordinates}

An alternative type of invariant, not necessarily a polynomial in the
coordinates of the state vector, is obtained by specifying a canonical
point on each orbit. The values of the invariant functions at any point
are then the coordinates of the canonical point on its orbit. The
canonical points lie on a manifold corresponding to the space of orbits,
and their coordinates can (at least locally) be expressed in terms of an
appropriate number of parameters. 

One form of canonical state was suggested independently by Linden and
Popescu \cite{NoahSandu} and by Schlienz \cite{Schlienz:PhD}, who
pointed out that any pure state of three qubits can be written as 
\be \begin{split} |\Psi\> = &\cos\theta|0\>\(\cos\phi|0\>|0\>+\sin\phi|1\>|1\>\)\\
      &+ \sin\theta|1\>\(r(-\sin\phi|0\>|0\> + \cos\phi|1\>|1\>) +
      s|0\>|1\> + t\e^{i\omega}|1\>|0\>\)
\end{split} \end{equation}
where $0\le\theta,\phi\le\pi/4,\;0\le\omega<2\pi$, and $r,s,t$ are
non-negative real numbers satisfying $r^2 + s^2 + t^2 = 1$.
Simpler canonical forms, in which the number of non-zero coefficients is
reduced to five, have since been proposed by Acin et al \cite{Acin} and
Carteret et al \cite{Schmidt}; the latter form is 
\[ 
p|100\> + q|010\> + r|001\> + s|111\> + t\e^{\i\theta}|000\>
\] 
where $p,q,r,s,t$ and $\theta$ are real parameters. It is straightforward to calculate the
invariants $I_1,\ldots,I_6$ in terms of either of the above sets of 
parameters; the results are not enlightening. 

We will now describe
another, more intrinsically defined, form of canonical point whose
coordinates are more simply related to $I_1,\ldots,I_6$. 

The three-particle state $|\Psi\>$ has three Schmidt decompositions:
\be \label{eq:Schmidt} \begin{split}
|\Psi\> &= \sum_i\alpha_i|\phi_i\>_A|\Phi_i\>_{BC}\\
         &= \sum_i\beta_i|\theta_i\>_B|\Theta_i\>_{AC}\\
         &= \sum_i\gamma_i|\chi_i\>_C|X_i\>_{AB}
\end{split} \end{equation}
where $\{|\phi_i\>\}$, $\{|\theta_i\>\}$ and $\{|\chi_i\>\}$ ($i=0,1$) are
orthonormal pairs of one-particle states, $\{|\Phi_i\>\}$, $\{|\Theta_i\>\}$ and
$\{|X_i\>\}$ are orthonormal pairs of two-particle states, the suffices
indicate which of the three particles $A, B, C$ are in which state, and
$\{\alpha_i\}$, $\{\beta_i\}$ and $\{\gamma_i\}$ are pairs of
non-negative real numbers satisfying 
\be \label{eq:I1again} \alpha_1^2 + \alpha_2^2 = \beta_1^2 + \beta_2^2 = 
   \gamma_1^2 + \gamma_2^2 = \<\Psi|\Psi\> = I_1.
\end{equation}
These Schmidt coefficients, being the positive square roots of the
eigenvalues of the one-particle density matrices $\rho_A, \rho_B,
\rho_C$, are related to the quartic invariants by
\be \label{eq:I234again} \begin{split}
  \alpha_1^4 + \alpha_2^4 &= \tr(\rho_A^2) = I_2,\\
  \beta_1^4 + \beta_2^4 &= \tr(\rho_B^2) = I_3,\\
  \gamma_1^4 + \gamma_2^4 &= \tr(\rho_C^2) = I_4. \end{split}
\end{equation} 
These equations have unique real non-negative solutions
for $\alpha_i, \beta_i, \gamma_i$ provided the invariants
$I_1,\ldots,I_4$ satisfy 
\[ I_1 > 0, \qquad \half I_1^2 \le I_2, I_3,
I_4 \le I_1^2. 
\] 
Now consider the coordinates $c^{ijk}$ of $|\Psi\>$
with respect to the canonical basis $|\phi_i\>_A |\theta_j\>_B
|\chi_k\>_C$. If the states $|\phi_i\>, |\theta_j\>, |\chi_k\>$ were uniquely
determined by $|\Psi\>$ --- and they almost are --- then the coordinates
$c^{ijk}$ would be local invariants. However, the Schmidt decompositions
do not determine the phases of $|\phi_i\>$, $|\theta_j\>$ and $|\chi_k\>$.
We can fix these by requiring that four of the $c^{ijk}$ should be real:
for example, we can change the phases of $|\phi_0\>$ and $|\phi_1\>$ to
make $c^{000}$ and $c^{100}$ real, then change the phases of
$|\theta_0\>$ and $|\theta_1\>$ to make $c^{001}$ and $c^{011}$ real,
simultaneously changing the phase of $|\chi_0\>$ to keep $c^{000}$ and
$c^{100}$ real. (It is easy to show that under the six-dimensional group
of phase changes of the basis vectors, the generic set of coordinates
has two-dimensional stabiliser, so that the orbits are four-dimensional
and therefore four phases can be removed.)

From the Schmidt decompositions we obtain the one-particle density
matrices
\be \begin{split} \rho_A &= \sum_i \alpha_i^2|\phi_i\>\<\phi_i|,\\
                 \rho_B &= \sum_i \beta_i^2|\theta_i\>\<\theta_i|,\\
                 \rho_C &= \sum_i \gamma_i^2|\chi_i\>\<\chi_i|.
\end{split} \end{equation}
Hence the coordinates $c^{ijk}$ satisfy
\be \label{ceqn} \begin{split} 
  \sum_{jk}c^{ijk}\overline{c}_{ljk} &= \alpha_i^2\delta^i_l,\\
  \sum_{ik}c^{ijk}\overline{c}_{imk} &= \beta_j^2\delta^j_m,\\
  \sum_{ij}c^{ijk}\overline{c}_{ijn} &= \gamma_k^2\delta^k_n.
\end{split} \end{equation}

To obtain a relation between the $c^{ijk}$ and Kempe's invariant $I_5$,
we calculate
\begin{align*} &\tr[(\rho_A\ox\rho_B)\rho_{AB}] \\
&= \tr\left[\(\sum_i\alpha_i^2|\phi_i\>_A\<\phi_i|_A\)
  \(\sum_j\beta_j^2|\theta_j\>_B\<\theta_j|_B\)
  \(\sum_k\gamma_k^2|X_k\>_{AB}\<X_k|_{AB}\)\right] \\
&=\sum_{ijk}\alpha_i^2\beta_j^2\gamma_k^2|\<\phi_i|\theta_j|X_k\>|^2.
\end{align*}
But 
\[ c^{ijk} = \<\phi_i|_A\<\theta_j|_B\<\chi_k|_C|\Psi\> 
  = \gamma_k\<\phi_i|_A\<\theta_j|_B|X_k\>_{AB}.
\]
Hence
\[ \tr(\rho_A\rho_B\rho_{AB}) = \sum_{ijk}\alpha_i^2\beta_j^2|c^{ijk}|^2
\]
and so, using \eqref{eq:I5},
\be \label{eq:I5again} I_5 = 3\sum_{ijk}\alpha_i^2\beta_j^2|c^{ijk}|^2 - \sum_i\alpha_i^6 -
\sum_j\beta_j^6. 
\end{equation}

Finally, the relation between the $c^{ijk}$ and the 3-tangle
$I_6$ needs a longer argument which we will not give here. The result
is 
\be \label{eq:I6again} I_6 = \det R \end{equation}
where
\[  R^i_j = (\alpha_i^4 + \alpha_i^2)\delta^i_j 
  - \sum_{kl}(\beta_k^2 + \gamma_l^2)c^{ikl}\overline{c}_{jkl}
\]

In order to determine how many states have the same values of the
invariants $I_1,\ldots,I_6$, and therefore how many further
discrete-valued invariants are needed to specify uniquely a pure state
of three qubits up to local transformations, one would need to find the
number of different sets of coordinates $c^{ijk}$ satisfying the reality
conditions given above and the equations \eqref{ceqn},
\eqref{eq:I5again} and \eqref{eq:I6again}, where $\alpha_i, \beta_i$ and
$\gamma_i$ are determined by \eqref{eq:I1again} and
\eqref{eq:I234again}.

\section*{Acknowledgements}

I am grateful to Bob Gingrich for finding an error in an earlier version of
this paper and for performing the calculations whose result is reported at the end of Section 3.

\providecommand{\bysame}{\leavevmode\hbox to3em{\hrulefill}\thinspace}

\end{document}